\documentclass[reprint,superscriptaddress,amsmath, amssymb,aps,prd,notitlepage,longbibliography,floatfix,nofootinbib,twocolumn]{revtex4-1}

\usepackage{amsmath}
\usepackage{lipsum}
\allowdisplaybreaks[4]
\usepackage{cancel}
\usepackage{extarrows}
\usepackage{tensor}     
\usepackage{float}
\usepackage[caption = false]{subfig} 
\usepackage[final]{graphicx}   
\usepackage[
colorlinks=true,        
citecolor=blue,         
linkcolor=blue,         
urlcolor=blue           
]{hyperref}             
\usepackage{bm}         
\usepackage{xcolor}     
\usepackage{lipsum}
\usepackage{color}      
\usepackage[utf8]{inputenc} 
\usepackage[section]{placeins} 
\usepackage{appendix}
\usepackage{units}
\usepackage[capitalise]{cleveref}
\usepackage{units}

\usepackage{booktabs}
\usepackage{multirow}
\usepackage{import}
\usepackage{adjustbox}
\usepackage{url}

\begin{document}

\title{Reconstruction of dark energy using DESI DR2}

\author{Xue Zhang}
\affiliation{Center for Gravitation and Cosmology, 
    College of Physical Science and Technology, \\
    Yangzhou University, Yangzhou 225009, China}
\author{Yin-Hao Xu}
\affiliation{Center for Gravitation and Cosmology, 
    College of Physical Science and Technology, \\
    Yangzhou University, Yangzhou 225009, China}
\author{Yu Sang}
\email{corresponding author: sangyu@yzu.edu.cn}
\affiliation{Center for Gravitation and Cosmology, 
    College of Physical Science and Technology, \\
    Yangzhou University, Yangzhou 225009, China}


\begin{abstract}
Using a model-independent Gaussian process (GP) method to reconstruct the dimensionless luminosity distance $D$ and its derivatives, we derive the evolution of the dimensionless Hubble parameter $E$, the deceleration parameter $q$, and the state parameter $w$ of dark energy. We utilize the PantheonPlus, SH0ES, and Gamma Ray Burst (GRB) data to derive the dimensionless luminosity distance $D$. Additionally, we employ observational $H(z)$ data (OHD) and baryon acoustic oscillations (BAO) from Dark Energy Spectroscopic Instrument (DESI) Data Release 2 (DR2) to obtain the first derivative of the dimensionless luminosity distance $D^{'}$. To obtain the reconstructed $D$ and $D^{'}$, we utilize the fiducial value from each dataset, with particular emphasis on the varying $H_0$. According to the reconstruction results obtained from PantheonPlus+SH0ES+GRB+OHD and PantheonPlus+SH0ES+GRB+OHD+DESI data, we find that $E$ are consistent with the predictions of the $\Lambda$CDM model at a $2\sigma$ confidence level within the redshift range of $z<2$. However, the reconstruction results for $q$ exhibit deviations from the $\Lambda$CDM model in the range of $z<0.3$. Furthermore, we observe that the mean value of $w$ exhibits evolving behavior, transiting from $w < -1$ to $w > -1$ around $z_{\rm wt}=0.464^{+0.235}_{-0.120}$. Combining data from DESI DR2 can slightly enhance the accuracy of our constraints.
\end{abstract}
\maketitle

\section{Introduction}
\label{intro}

The accelerating expansion of the Universe was first discovered through observations of type Ia supernovae \cite{SupernovaSearchTeam:1998fmf,SupernovaCosmologyProject:1998vns} in the late 20th century. This finding was subsequently confirmed by a variety of cosmological experiments, including studies on large-scale structure \cite{SDSS:2003eyi}, anisotropy of the cosmic microwave background \cite{WMAP:2003elm}, and measurements of baryon acoustic oscillations \cite{SDSS:2005xqv}. 
To explain the phenomenon of accelerated expansion, numerous theoretical frameworks have been proposed, among which dark energy theory has attracted significant attention. One of the most prominent candidates within dark energy theories is the $\Lambda$CDM model. This model exhibits strong compatibility with numerous observational datasets and experimental results. However, it continues to confront two major challenges known as the fine-tuning problem and coincidence problem \cite{Weinberg:1988cp}. Therefore, it remains essential to continually utilize observational data to test and refine the $\Lambda$CDM model. 
The evolution of the Universe is typically characterized by several key parameters: the dimensionless Hubble parameter $E$, the deceleration parameter $q$, and the state parameter $w$ of dark energy. The sign (positive or negative) of the deceleration parameter indicates whether the expansion of the universe is decelerating or accelerating. Meanwhile, the state parameter $w$, defined as the ratio of pressure to energy density, takes a value of $w=-1$ within the framework of $\Lambda$CDM model.

The Gaussian process is a widely utilized statistical numerical method in cosmology. This model-independent method does not depend on any specific theoretical model and can directly reconstruct the parameters from observational datasets. The only assumption is that the reconstructed parameters follow a Gaussian distribution. Due to the model-independent characteristics of the reconstruction results, we can compare these results with those derived from the $\Lambda$CDM model for validation purposes. 
The Gaussian processes method has been extensively employed across various fields of cosmology, including the reconstruction of dark energy \cite{Seikel:2012uu,Zhang:2018gjb,Wang:2017jdm,Wang:2019ufm,Lin:2019cuy,Wang:2022xdw,Velazquez:2024aya,Holsclaw:2010sk,Ghosh:2024kyd,Dinda:2024xla,Bonilla:2020wbn}, testing the validity of the concordance model of cosmology \cite{Dialektopoulos:2023dhb}, examining the cosmic distance duality relation \cite{Mukherjee:2021kcu}, constraining spatial curvature \cite{Liu:2024yib,Wang:2022rvf,Ran:2023jmh,Liu:2020pfa}, exploring interaction between dark energy and dark matter \cite{Yang:2015tzc,Mukherjee:2021ggf,Cai:2017yww,Bonilla:2021dql,Escamilla:2023shf,vonMarttens:2020apn,Bonilla:2021dql}, conducting null test of dynamical dark energy \cite{Dialektopoulos:2023dhb,Dinda:2024kjf}, determining the characteristic length scale $r_{\rm s}$ of the baryon acoustic oscillations \cite{Lemos:2023qoy}, reconstructing the scalar field potential for dark energy \cite{Niu:2023hak,Piloyan:2018xwa}, reconstructing modified gravity (such as $f(Q)$ gravity \cite{Gadbail:2024lek,Gadbail:2024rpp,Yang:2024tkw}, $f(T)$ gravity \cite{Cai:2019bdh,Ren:2022aeo}, $f(R,T)$ gravity \cite{Bernardo:2021qhu} and Horndeski gravity \cite{Bernardo:2021qhu}), governing the
evolution of the temperature of the cosmic microwave background radiation (CMB) \cite{Avila:2025dne}, measuring the growth index $\gamma$, growth rate $f$ and $f \sigma_8$ \cite{Oliveira:2023uid,Avila:2022xad}, resolving the Hubble tension through observational data \cite{Gomez-Valent:2023uof,Raveri:2023zmr,DAgostino:2023cgx,Yang:2022jkf,Abdalla:2022yfr,CosmoVerseNetwork:2025alb}, and so on.

In the context of investigating the properties of dark energy, we present some relevant work as follows.
Holsclaw et al. reconstructed the redshift evolution of the equation of state parameter $w$ using a nonparametric method based on Gaussian process modeling and Markov chain Monte Carlo sampling \cite{Holsclaw:2010sk}. Seikel et al. published the GAPP code, a program designed to reconstruct dark energy and expansion dynamics through Gaussian processes, employing SNIa Union2.1 data and the mock DES data to effectively reconstruct the state parameters $w$ \cite{Seikel:2012uu}. Yang et al. reconstructed the interaction between dark energy and dark matter utilizing SNIa Union 2.1 data \cite{Yang:2015tzc}. 
Wang et al. used a combination of the Union 2.1 SNIa data, cosmic chronometer $H(z)$ data, and Planck’s shift parameter within the Gaussian processes method to explore how various matter density parameters $\Omega_{\rm m}$, curvature parameters $\Omega_{\rm k}$, and Hubble parameters $H_{0}$ influence reconstruction results \cite{Wang:2017jdm}. Both the background datasets, including supernova and $H(z)$ data, along with perturbation data from the growth rate indicated a possible existence of dynamic dark energy \cite{Zhang:2018gjb}. 
Lin et al. combined the Pantheon dataset with the $H(z)$ dataset, inferring that $H_0=70.5\pm0.5~\mathrm{km\ s^{-1} Mpc^{-1}}$ without imposing any prior on $H_0$ \cite{Lin:2019cuy}. This result has helped alleviate the tension between locally measured values of $H_{0}$ and those measured globally. Recently, Ghosh et al. reconstructed dimensionless Hubble parameters $H(z)$ and deceleration parameters $q$ utilizing data from DESI DR1 and SDSS \cite{Ghosh:2024kyd}. Their findings revealed a significant discrepancy in the reconstruction of $H(z)$ and $q$ when using DESI DR1 or SDSS independently. However, the combined analysis of DESI DR1 and SDSS data produced results that are consistent with the $\Lambda$CDM model. 
Some very recent works \cite{Pang:2024qyh,Nagpal:2025omq,Chaudhary:2025pcc,Sharma:2025qmv} further explore model-independent reconstructions and evolving dark energy using DESI DR2 data.

In this paper, we utilize observational data from PantheonPlus+SH0SE \cite{Brout:2022vxf}, Gamma Ray Bursts (GRB) \cite{Mu:2023bsf},  observational $H(z)$ data (OHD) \cite{Qi:2023oxv}, and DESI DR2 BAO \cite{DESI:2025zgx}. These observation datasets serve as priors for our Gaussian process. We begin with PantheonPlus+SH0SE as the reconstructed basic observational data and sequentially incorporate the GRB, OHD, and DESI data as joint datasets to explore their impact on Gaussian process reconstruction while comparing these reconstruction results with those derived from the $\Lambda$CDM model. We aim to enhance the accuracy of reconstruction results at high redshift by including GRB data as supplementary sources. By including OHD and DESI data, we aim to enhance the reconstruction results of $D^{'}$ and $D^{''}$, thereby improving the estimates of $E$, $q$, and $w$. 
In contrast to previous studies \cite{Zhang:2018gjb,Lin:2019cuy,Wang:2022xdw}, we do not use a specific value of $H_{0}$ as a prior for the entire reconstruction process. It is important to note that the supernova, GRB, OHD, and DESI data are independent of each other and are derived from different fiducial values of $H_{0}$. Therefore, using an uniform $H_{0}$ as a prior for joint datasets reconstruction is not reasonable. So we adopt the value of $H_{0}=73.6\pm1.1~\mathrm{km\ s^{-1} Mpc^{-1}}$ derived from the PantheonPlus+SH0SE \cite{Brout:2022vxf} cosmological parameter constraints in the $\Lambda$CDM model as our reconstruction prior for both PantheonPlus+SH0SE and GRB datasets. For OHD and DESI, we utilize the value of 
$H_{0}=68.17\pm0.28~\mathrm{km\ s^{-1} Mpc^{-1}}$ 
provided by DESI DR2 cosmological parameter constraints in the $\Lambda$CDM model as our reconstruction prior.

In the following sections, we introduce the methodology and Gaussian process reconstruction method employed in this paper in Sect. \ref{sec:meth}, present the observation datasets we used in Sect. \ref{sec:data}, give a discussion of the reconstruction results in Sect. \ref{sec:res}, and summarize the reconstruction results in Sect. \ref{sec:d and c}.

\section{Methodology}
\label{sec:meth}
In this section, we present the theoretical basis of this paper and describe the Gaussian Process methodology employed for reconstruction.
\subsection{Theoretical basis}
\label{subsec:theoretical}
In the Friedmann–Robertson–Walker (FRW) universe, the luminosity distance $d_{L}(z)$ of SN Ia is represented as
\begin{equation}\label{eq:dl}
d_{L}(z) = \frac{c}{H_{0}}(1+z)\int_{0}^{z}\frac{H_{0}}{H(z^{'})}{\rm d}z^{'}, 
\end{equation}
while the dimensionless comoving luminosity distance $D$ can be defined using 
\begin{equation}\label{eq:dz}
D\equiv\frac{H_{0}}{c}\frac{d_{L}(z)}{1+z}.
\end{equation}

By combining Eqs. (\ref{eq:dl}) and (\ref{eq:dz}) and taking the derivative with respect to redshifts $z$, we can obtain the relationship between the Hubble parameter and the dimensionless luminosity distance. Moreover, the dimensionless Hubble parameter $E(z)$ can be defined as
\begin{equation}\label{3}
E(z)\equiv\frac{H(z)}{H_{0}}=\frac{1}{D^{'}},
\end{equation}
where the superscript prime denotes the derivative with respect to the redshift $z$.

We consider a flat FRW universe with dark matter and dark energy, where in the evolution is governed by the Friedman equation
\begin{equation}
E(z)=\sqrt{\Omega_{\rm m}(1+z)^{3}+\Omega_{\rm de}\exp\left[3\int_{0}^{z}\frac{1+w(z^{'})}{1+z^{'}}{\rm d}z^{'}\right] },
    \label{4}
\end{equation}
where $\Omega_{\rm de}=1-\Omega_{\rm m}$.
Then, using Eqs. (\ref{3}) and (\ref{4}), the state equation of dark energy can be obtained as
\begin{equation}\label{5}
w=\frac{1}{3}\frac{-2(1+z)D^{''}-3D^{'}}{D^{'}-\Omega_{\rm m}(1+z)^{3}D^{'3}}.    
\end{equation}
The acceleration of the Universe's expansion is typically quantified by the deceleration parameter. It is defined as $q=-\ddot{a}a/(\dot{a}^{2})$, where $a=1/(1+z)$ represents the scale factor for the expansion of the universe, and the dot indicates a derivative with respect to cosmic time.
We can obtain a relationship between the deceleration parameter and $D^{'}$, $D^{''}$ as follows: 
\begin{equation}\label{6}
q=(1+z)\frac{H^{'}}{H}-1=-(1+z)\frac{D^{''}}{D^{'}}-1.
\end{equation}
From Eqs. (\ref{3}), (\ref{5}), and (\ref{6}), it is evident that the cosmological parameter $E$, $w$, and $q$ can be derived by $D^{'}$ and $D^{''}$.
In Sect. \ref{sec:data}, we will describe the observational data used in the reconstruction of cosmological parameters, including SNIa, GRB, OHD and BAO. Specifically, we will discuss how to obtain the $D$ and its derivatives required by the GP from the observed data.

\subsection{Gaussian Processes}\label{subsec:GP}
Gaussian Process is a model-independent method that can be applied to any parameterization. 
For example, it can be used for testing the geometric parameterization models (see Ref. \cite{Pacif:2020hai} for the geometric parametric models), such as $q(z)$, $H(z)$, $a(t)$, or $j(z)$ models.
In essence, a Gaussian process is an infinite-dimensional Gaussian distribution that describes the distribution of functions, while a Gaussian distribution describes the distribution of random variables. When reconstructing the objective function through Gaussian process using observation data $D=\{  (x_{i},y_{i})|i=1,...,n\}$, it suffices to assume that each set of observation data follows a Gaussian distribution, i.e.,
\begin{equation}\label{7}
\boldsymbol{y}\sim N\left(\bm{\mu}, \mathbf{K}\boldsymbol{(x, x)}+\mathbf{C}\right),  
\end{equation}
where $\boldsymbol{x}=\{ x_{i}\}$, $\bm{\mu}$ represents the mean of the Gaussian process, $\mathbf{C}$ denotes the covariance matrix of the data, and $\mathbf{K}\boldsymbol{(x,x)}$ is the covariance matrix provided by the covariance function. 

The posterior distribution of the function $f^{*}=f(\boldsymbol{x}^{*})$ ($\boldsymbol{x}^{*}=\{x^{*}_{i}\}$ represents the position where the function needs to be reconstructed) that we aim to reconstruct can be represented by a joint Gaussian distribution of different observation data. Therefore, it is crucial to select appropriate covariance and mean functions in Gaussian process reconstruction. Even with the same observation data, choosing different covariance functions and mean functions will yield different reconstruction results. 
The Ref. \cite{Seikel:2013fda} compares four kernels: the squared exponential and three members of the Matérn class ($\nu$=5/2,7/2,9/2). The Matérn class provide a flexible family where the smoothness can be tuned via the parameter $\nu$. The squared exponential should be used with caution, as it often underestimates uncertainty. For smooth models (e.g., $\Lambda$CDM model), Matérn(9/2) is recommended. For models with more structure, Matérn(7/2) may be more reliable. Matérn(5/2) is generally too conservative. Thus, the kernel choice should align with the expected smoothness of the underlying dark energy dynamics. 
In this paper, we adopt the Mat\'{e}rn ($v$=9/2) covariance function for Gaussian process reconstruction based on the analysis in Ref. \cite{Seikel:2013fda}, which indicates that Mat\'{e}rn ($v$=9/2) leads to more stable results and a smoother reconstructed image. The covariance function is defined as
\begin{align}\nonumber
k(x,\tilde{x})=&\sigma^{2}_{f}\exp\left(-\frac{3|x-\tilde{x}|}{l}\right)
\bigg[1+\frac{3|x-\tilde{x}|}{l}\\
&+\frac{27(x-\tilde{x})^{2}}{7l^{2}}
+\frac{18|x-\tilde{x}|^{3}}{7l^{3}}
+\frac{27(x-\tilde{x})}{35l^{4}}\bigg],
\end{align}
where $\sigma_{f}$ and $l$ are hyperparameters that can be optimized by maximizing marginalized likelihood. 
To reconstruct the dimensionless Hubble parameter $E$, the deceleration parameters $q$, and the equation of state $w$, we modified the package GaPP3 (Gaussian Processes in Python3)\footnote{https://github.com/lighink/GaPP3}. 
See Ref. \cite{Seikel:2012uu} for more details. 

\section{Data}\label{sec:data}
In this paper, we utilized observational data for GP reconstruction, including a type Ia supernovae sample (SNe Ia), SH0ES, GRB, OHD and DESI. GP reconstruction performs well in reconstructing the objective function. However, when attempting to simultaneously reconstruct the first and second derivatives of the objective function, the error of the observed data will be amplified. By incorporating both the objective function and its derivative as priors information for reconstruction, the accuracy of the reconstruction can be substantially improved. In terms of the type of data ultimately input into GP reconstruction, the observation data used in this article can be categorized into two types: one derived from calculations to obtain $D$, and the other from calculations to obtain $D^{'}$. We set the boundary conditions as $D(z=0)=0$ and $D^{'}(z=0)=1$ for Gaussian process reconstruction.

\subsection{Type Ia supernovae}
\label{subsec:supernova}

SNe Ia, due to their unique formation process, exhibit the same absolute magnitude. This characteristic makes them reliable standard candles for cosmological measurements. For our analysis, we use the Pantheon+ data \cite{Brout:2022vxf}. In the following context, we will denote the Pantheon+ dataset as PantheonPlus. The PantheonPlus analysis proposed constraints on cosmological parameters, including distance modulus and its error. The dataset comprises 1701 light curves of 1550 distinct Type Ia supernovae in redshift range $0.001<z<2.26$. We utilize the distance modulus and corresponding redshift calibrated with the Cepheid variable provided by SH0ES, including their associated uncertainties, and refer to this dataset as PantheonPlus+SH0ES. Compared to earlier Type Ia supernovae observation datasets such as Union2.1 \cite{SupernovaCosmologyProject:2011ycw}, JLA \cite{SDSS:2014iwm} and Pantheon \cite{Pan-STARRS1:2017jku}, PantheonPlus + SH0ES contains a larger number of data points and exhibits smaller uncertainties, leading to better GP reconstruction results.

There exists a relationship between the distance modulus of Type Ia supernovae and their corresponding luminosity distance, expressed as
\begin{equation}\label{9}
\mu(z)=5\log_{10}\frac{d_{L}(z)}{\rm Mpc}+25,
\end{equation}
where $\mu$ is the distance modulus of a Type Ia supernovae, which is dimensionless. 
By substituting Eq. (\ref{eq:dz}) into (\ref{9}), 
we can derive the relationship between the dimensionless luminosity distance and distance modulus given by
\begin{equation}\label{10}
D=\frac{H_{0}~\rm Mpc}{c}\frac{10^{\frac{\mu-25}{5}}}{1+z}.    
\end{equation}

We utilize the value of $H_{0}=73.6\pm1.1~\mathrm{km\ s^{-1} Mpc^{-1}}$ as a prior for reconstructing with PantheonPlus+SH0SE data \cite{Brout:2022vxf}. The total error matrix ($\bm{\Sigma}_{\mu}$) for $\mu$ is comprised of the systematic error matrix ($\mathbf C_{\rm sys}$) and the statistical error matrix ($\mathbf C_{\rm stat}$), which can be expressed as follows:
\begin{equation}\label{11}
\bm{\Sigma}_{\rm\mu}=\mathbf C_{\rm sys}+\mathbf C_{\rm stat}.  \end{equation}
The error matrix of $D$ can be derived using the standard error propagation formula, given as
\begin{equation}\label{12}
\bm{\Sigma}_{\rm Dz}=\mathbf D_{1}\bm{\Sigma}_{\rm\mu}\mathbf D^{T}_{1},   
\end{equation}
Here, $\bm{\Sigma}_{\rm Dz}$ denotes the error matrix of $D$. The superscript ‘T’ indicates the transpose of the matrix, and $\mathbf D_{1}$ is defined as the Jacobian matrix:
\begin{equation}
\mathbf D_{1}= {\rm diag}(\frac{\ln{10}}{5} \mathbf{Dz}).    
\end{equation}
In this context, $\mathbf{Dz}$ represents a vector whose components are the dimensionless luminosity distances of all Type Ia supernovae (1701 data points).

\subsection{Gamma Ray Bursts}
\label{subsec:GRB}

We utilize a dataset comprising 97 data points, which include redshift, distance modulus, and distance modulus error as provided in Ref. \cite{Mu:2023bsf}. We use Pantheon+ samples \cite{Brout:2022vxf} to update the distance modulus of GRB employing the method proposed in Ref. \cite{Liang:2022smf}. The analysis includes a total of 182 datasets, with a redshift range from $0.8<z \leq 8.2$.
However, not all of these 182 datasets possess physical significance. Therefore, we applied the screening method described in Ref. \cite{Mu:2023bsf} to identify and retain 97 datasets with physically meaningful. The GRB data exhibit higher redshifts compared to those from PantheonPlus+SH0SE, with the maximum redshift reaching $z=8.2$. We aim to enhance the accuracy of our reconstructed results within high-redshift phases by incorporating GRB data into our analysis.

The data provided by GRB includes redshift, distance modulus and the error of distance modulus, denoted as $(z_{\rm GRB},~\mu_{\rm GRB},~\sigma_{\mu_{\rm GRB}})$. Consequently, we can derive the dimensionless luminosity distance $D$ using Eq. \eqref{10}. 
To calculate $\sigma_{{\rm D}z_{\rm GRB}}$, we employ the the following formula:
\begin{equation}
\sigma^{2}_{{\rm D}z_{\rm GRB}}=(\frac{\partial {\rm D}z_{\rm GRB}}{\partial \mu_{\rm GRB}})^{2}\sigma^{2}_{\mu_{\rm GRB}}.    
\end{equation}

\subsection{observational $H(z)$ data}
\label{subsec:H(z)}
 
The OHD is primarily obtained using two methods. The first method involves a cosmic chronometer (CC) \cite{Jimenez:2001gg,Simon:2004tf,Stern:2009ep}, which estimates the ages of various galaxies in the universe to derive $H(z)$. The second method is based on the BAO peak in the galaxy power spectrum \cite{Gaztanaga:2008xz,Moresco:2012jh} or utilizes the BAO peak from the Ly$\alpha$ forest of quasars. This article utilizes the data presented TABLE I and TABLE II from Ref. \cite{Qi:2023oxv}. The information in these tables contain Hubble parameters corresponding to redshift, along with their corresponding errors. Specifically, the data in TABLE I is obtained through cosmic chronometeor, thus referred to as ``CC $H(z)$'' in this paper. On the other hand, the data in TABLE II is inferred from the peak of BAO observed in the galaxy power spectrum. Therefore, we denote it as ``BAO $H(z)$''.

By slightly modifying Eq. \eqref{3}, we can derive
\begin{equation}
D^{'}=\frac{H_{0}}{H(z)}.    
\end{equation}
We adopt the value of 
$H_{0}=68.17\pm0.28~\mathrm{km\ s^{-1} Mpc^{-1}}$ 
\cite{DESI:2025zgx} as a prior for OHD measurements. The uncertainty in $D^{'}$ is calculated using the standard error propagation formula
\begin{equation}
\sigma^{2}_{D^{'}}=(\frac{\partial D^{'}}{\partial H})^{2}
\sigma^{2}_{H}. 
\end{equation}

\subsection{DESI DR2 BAO}
\label{subsec:DESI}

The detailed data of DESI DR2 BAO is provided in Table IV of Ref. \cite{DESI:2025zgx}. 
The data provided by DESI is categorized into three distinct forms:
\begin{itemize}
    \item 
    $D_{\rm H}/r_{\rm d}$: Here, $r_{\rm d}$ denotes the sound horizon at the drag epoch, while $D_{\rm H}$ is known as equivalent distance variable and has a defined relationship with the Hubble parameters:
    \begin{equation}
    D_{\rm H}(z)=\frac{c}{H(z)}.    
    \end{equation}
    \item 
    $D_{\rm M}/r_{\rm d}$: In this case, $D_{\rm M}(z)$ refers to the comoving distance. The corresponding relationship is given by:
    \begin{equation}
    D_{\rm M}(z)=\int^{z}_{0}D_{\rm H}(z^{'}){\rm d} z^{'}.    
    \end{equation}
    \item   
    $D_{\rm V}/r_{\rm d}$: Here, $D_{\rm V}(z)$ represents the angular-average distance. The relationship can be expressed as:
    \begin{equation}
    D_{\rm V}(z)=\left[zD^{2}_{\rm M}(z)D_{\rm H}(z)\right]^{1/3}.    
    \end{equation}
\end{itemize}

Due to $D_{\rm M}(z)$ and $D_{\rm V}(z)$ are related to the integral of $D^{'}$, they are unable to be used in the construction. Therefore, we utilize the DESI DR2 observation data only in the form of ratio $D_{\rm H}/r_{\rm d}$. This includes luminous red galaxies (LRG), emission line galaxies (ELG), quasars (QSO) and Lyman-$\alpha$ forest (Ly$\alpha$), with effective redshifts of $z_{\rm eff}=(0.510, 0.706, 0.934, 1.321, 1.484, 2.330)$. In this article, we adopt the value $r_{\rm d}=147.05\pm0.30~{\rm Mpc}$ as reported by Planck 2018 \cite{Planck:2018vyg}. We derive $H(z)$ using the following equation:
\begin{equation}
H(z)=\frac{c}{D_{\rm H}/r_{\rm d}\times r_{\rm d}}.    
\end{equation}
To obtain the error of $H(z)$, we apply  the error propagation formula:
\begin{equation}
\sigma^{2}_{{H}}=\left[\frac{\partial H}{\partial (D_{\rm H}/r_{\rm d})}\right]^{2}
\sigma^{2}_{D_{\rm H}/r_{\rm d}}
+\left[\frac{\partial H}{\partial r_{\rm d}}\right]^{2}
\sigma^{2}_{r_{\rm d}}.   
\end{equation}

\section{Result}\label{sec:res}
In this section, we present the results of reconstructing the parameters $D$, $E$, $q$, $w$ utilizing joint datasets from various observational data.

\subsection{The reconstruction of $D$}
\label{subsec:rec D}

\begin{figure*}[!htbp]
  \centering
  \subfloat{\includegraphics[width=0.99\linewidth]{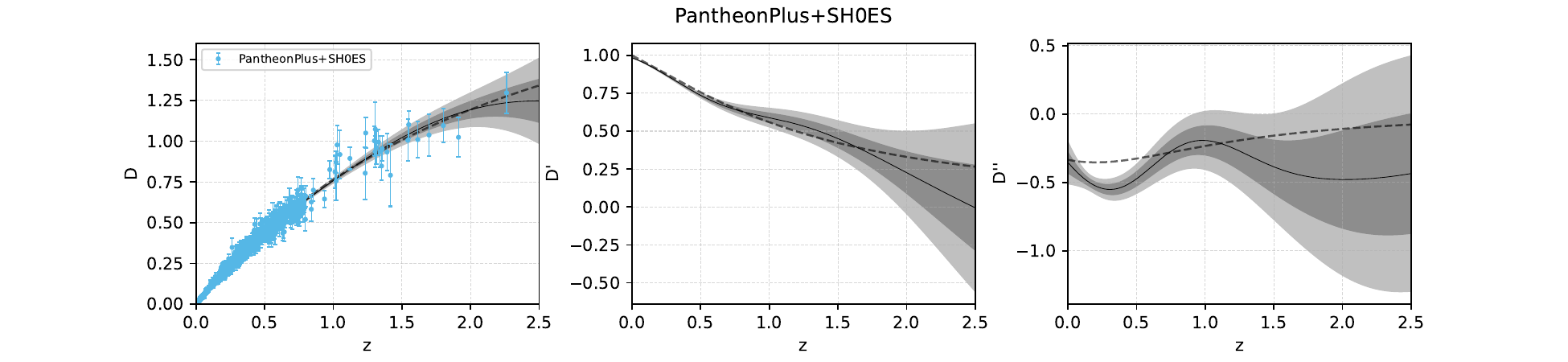}}\\
  \subfloat{\includegraphics[width=0.99\linewidth]{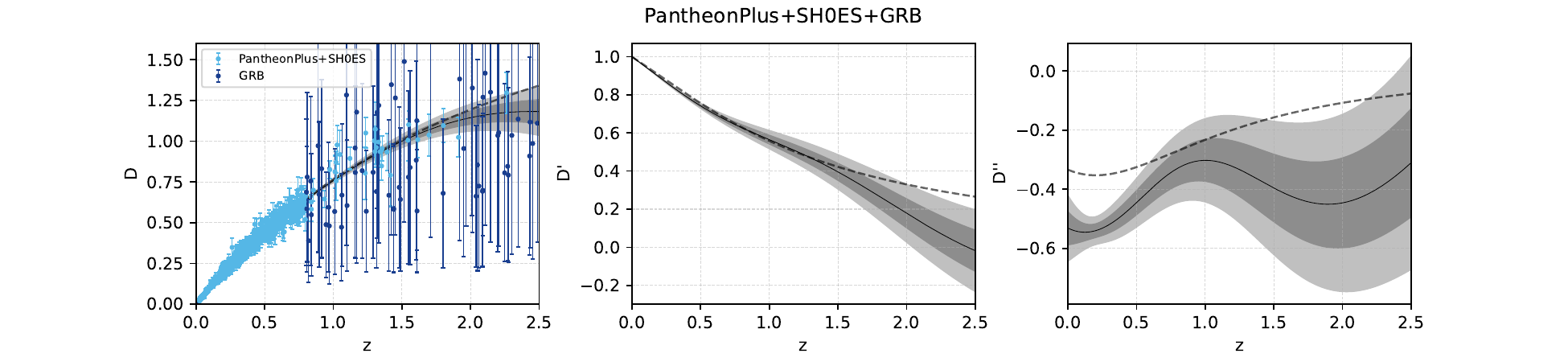}}\\
  \subfloat{\includegraphics[width=0.99\linewidth]{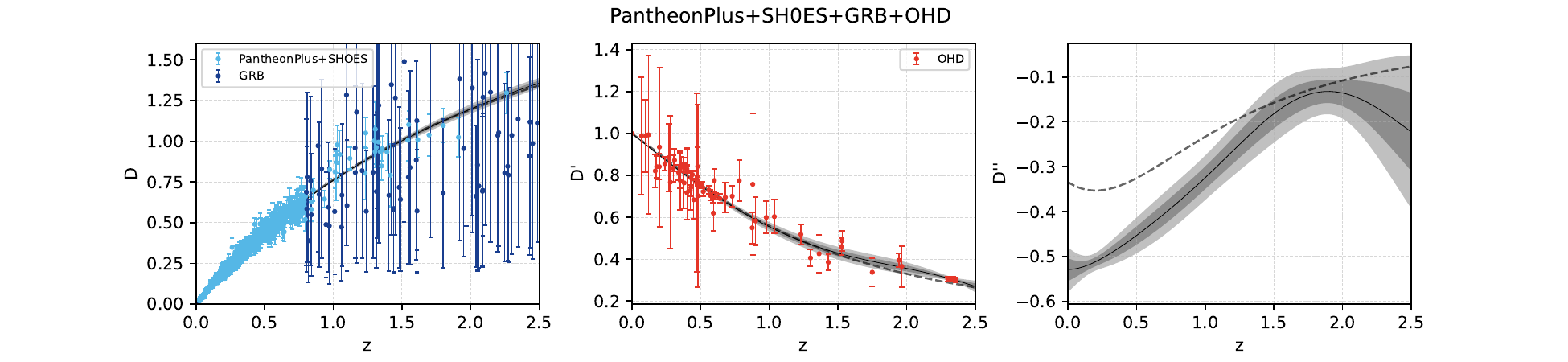}}\\
  \subfloat{\includegraphics[width=0.99\linewidth]{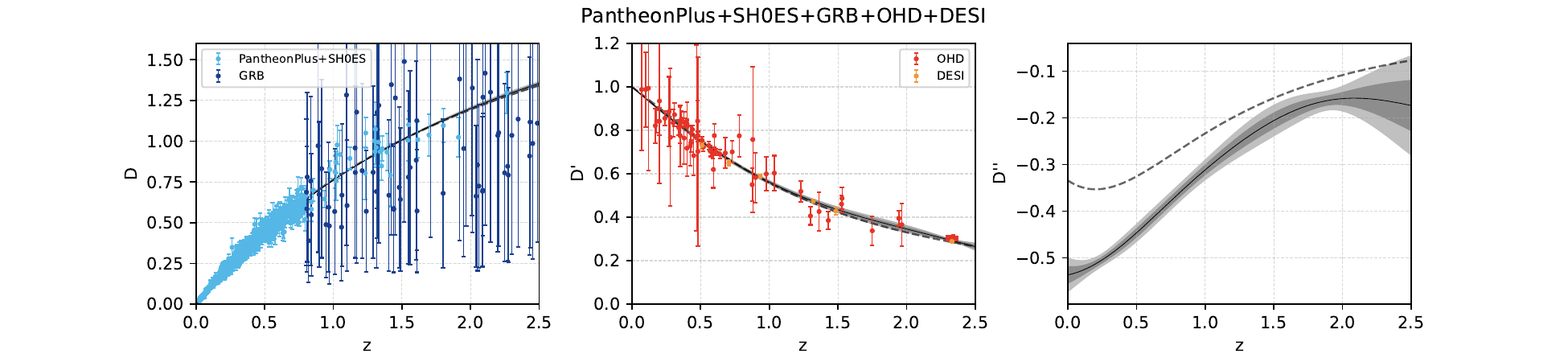}}
  \caption{The reconstruction of $D$ along with its first and second derivatives. It is organized into four rows, corresponding to four different joint datasets: PantheonPlus+SH0ES, PantheonPlus+SH0ES+GRB, PantheonPlus+SH0ES+GRB+OHD, and PantheonPlus+SH0ES+GRB+OHD+DESI.
  The black dashed line represents the theoretical curve of the $\Lambda$CDM model. The shaded areas in dark grey and light grey are the confidence intervals of 1$\sigma$ and 2$\sigma$, respectively. Additionally, the mean values and error bars, depicted in light blue, dark blue, red and orange, correspond to the PantheonPlus+SH0ES, GRB, OHD and DESI datasets, respectively.}
  \label{fig_D}
\end{figure*}

In Fig.~\ref{fig_D}, we present the dimensionless luminosity distance $D$ along with its first and second derivatives, which are reconstructed from a combination of various observation datasets. 

The results of PantheonPlus+SH0ES reconstruction for $D$ and its first derivatives agree well with the curve of $\Lambda$CDM model\footnote{Here, we adopt the parameter values from Planck 2018 \cite{Planck:2018vyg}: $H_{0}=67.4\pm0.5\rm km\cdot\rm s^{-1}\cdot\rm Mpc^{-1}$, $\Omega_{\rm m0}=0.334\pm 0.018$. } (the black dashed line)  within a range of 2$\sigma$. However, due to insufficient observation data at high redshifts ($z>1.5$) in these samples, the reconstruction results exhibit relatively large errors at high redshifts. After including the GRB data, there has been a significant improvement in accuracy at high redshift. The reconstruction results for $D$ are consistent with the $\Lambda$CDM model at a confidence interval of 2$\sigma$. However, the reconstruction results for $D^{'}$ exhibit a deviation from the $\Lambda$CDM model at high redshifts ($z>2$). 

As illustrated in the third row of Fig. \ref{fig_D}, the reconstruction error of $D$ is significantly reduced after the incorporation of OHD, which is associated with the derivative data $D^{'}$. This finding suggests that adding derivative data of the function as a Gaussian process prior during its reconstruction can indeed enhance the quality of our results. 
Furthermore, after incorporating OHD, the deviation trend of the $\Lambda$CDM model in the result of PantheonPlus+SH0ES+GRB reconstruction returns to within the 2$\sigma$ confidence level. As shown in the bottom row figure of Fig. \ref{fig_D}, the inclusion of DESI data further reduces the error in the reconstructed result of $D$ and its derivatives. 
So does the reconstruction for $D^{''}$, although the center values are deviating from the $\Lambda$CDM model.

\subsection{The reconstruction of $E$}
\label{subsec:rec E}

\begin{figure*}[!htbp]
  \centering
  \subfloat{\includegraphics[width=0.33\linewidth]{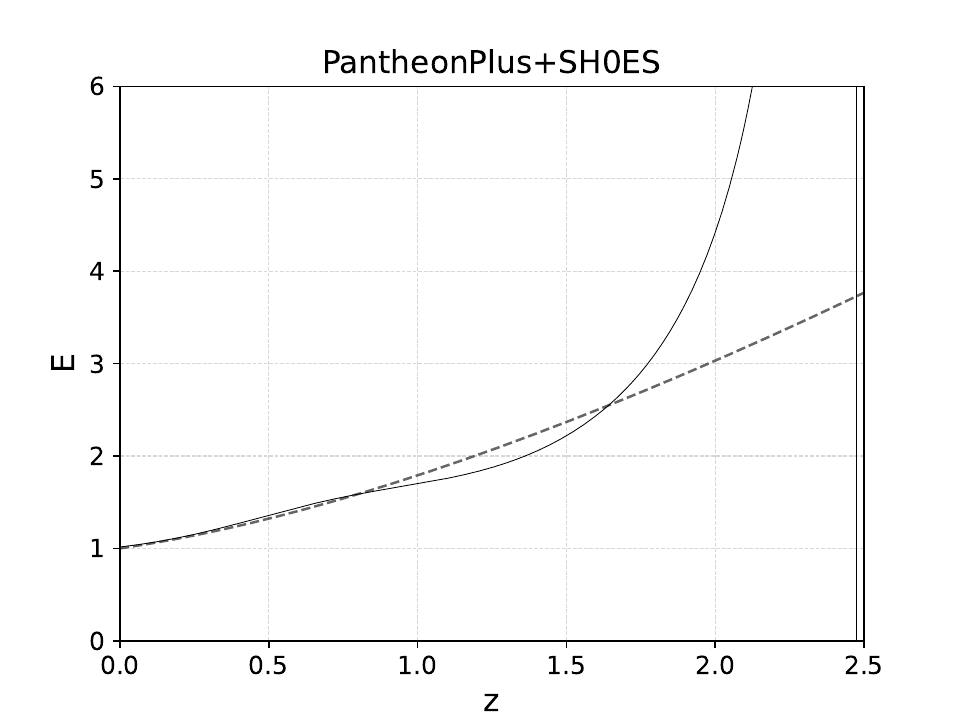}}
  \subfloat{\includegraphics[width=0.33\linewidth]{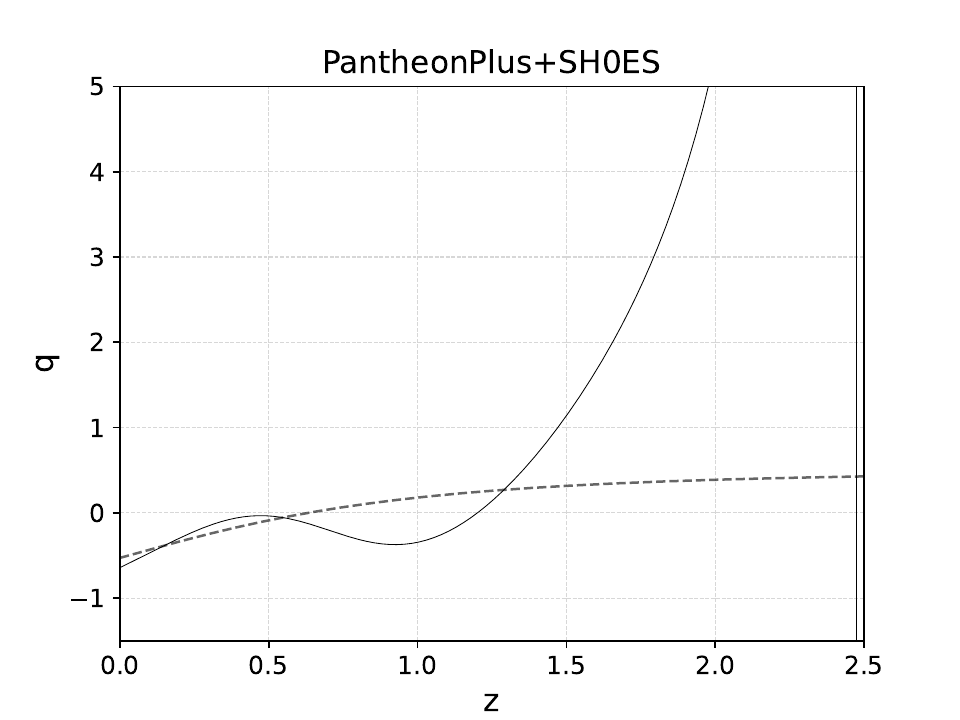}}
  \subfloat{\includegraphics[width=0.32\linewidth]{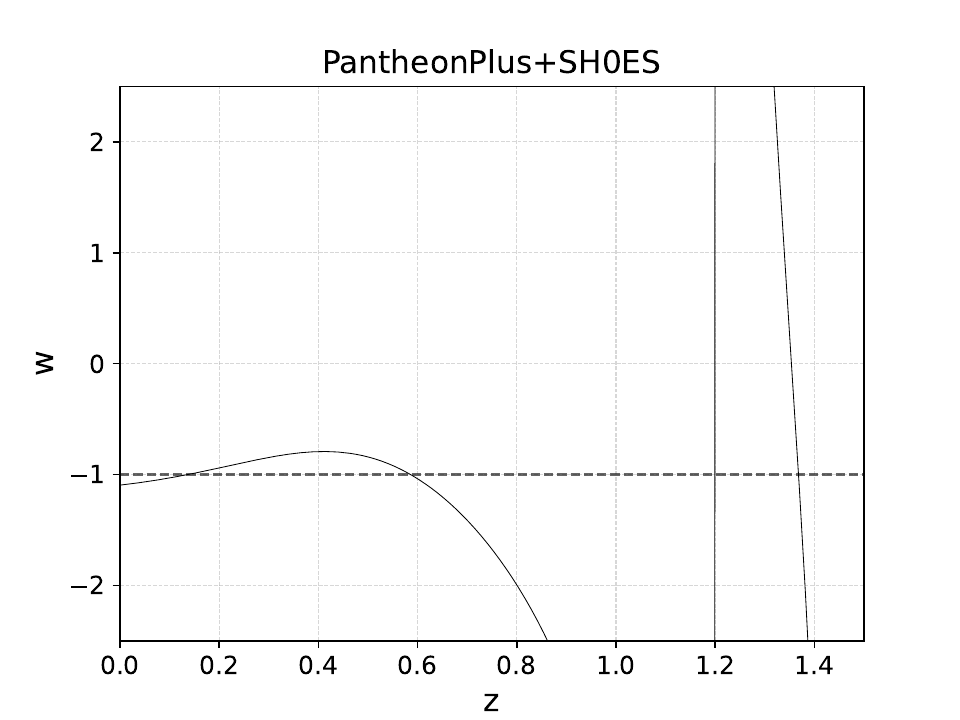}}\\
  \subfloat{\includegraphics[width=0.32\linewidth]{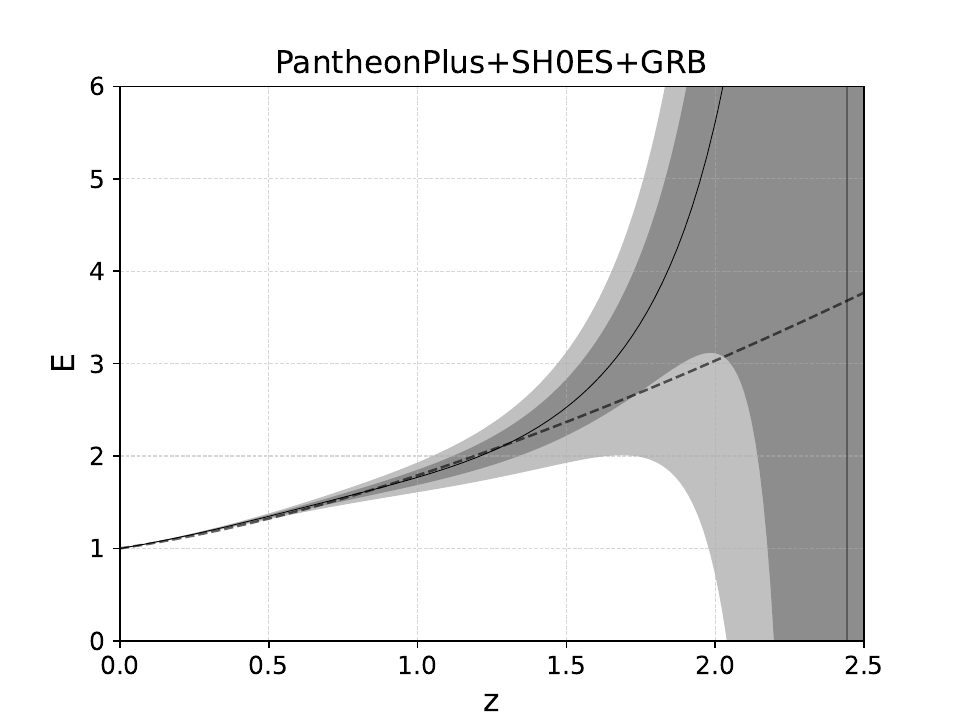}}
  \subfloat{\includegraphics[width=0.32\linewidth]{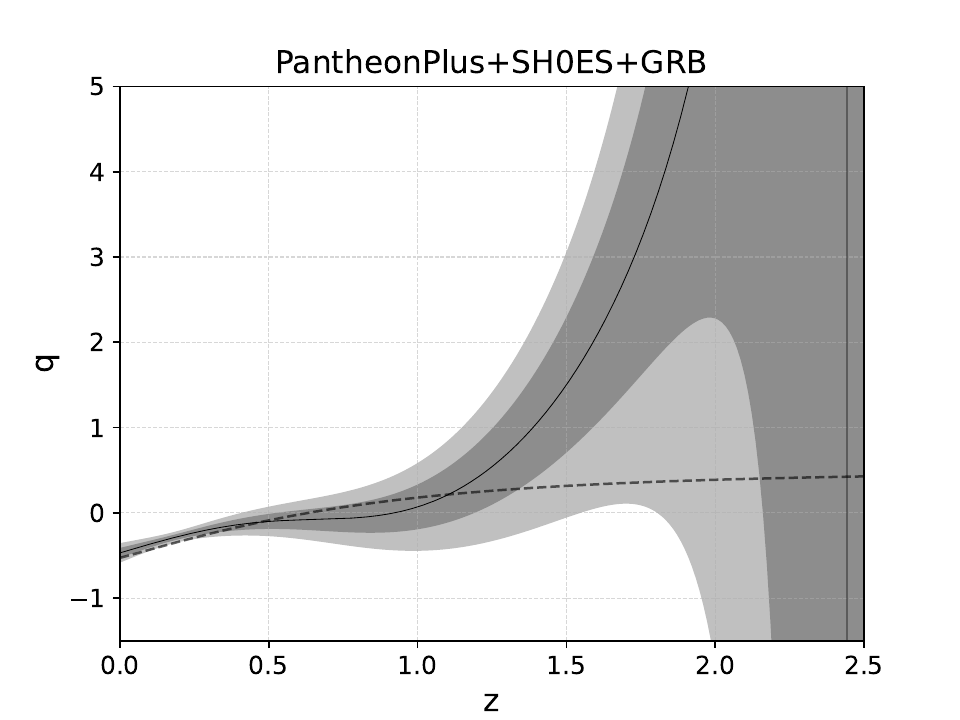}}
  \subfloat{\includegraphics[width=0.32\linewidth]{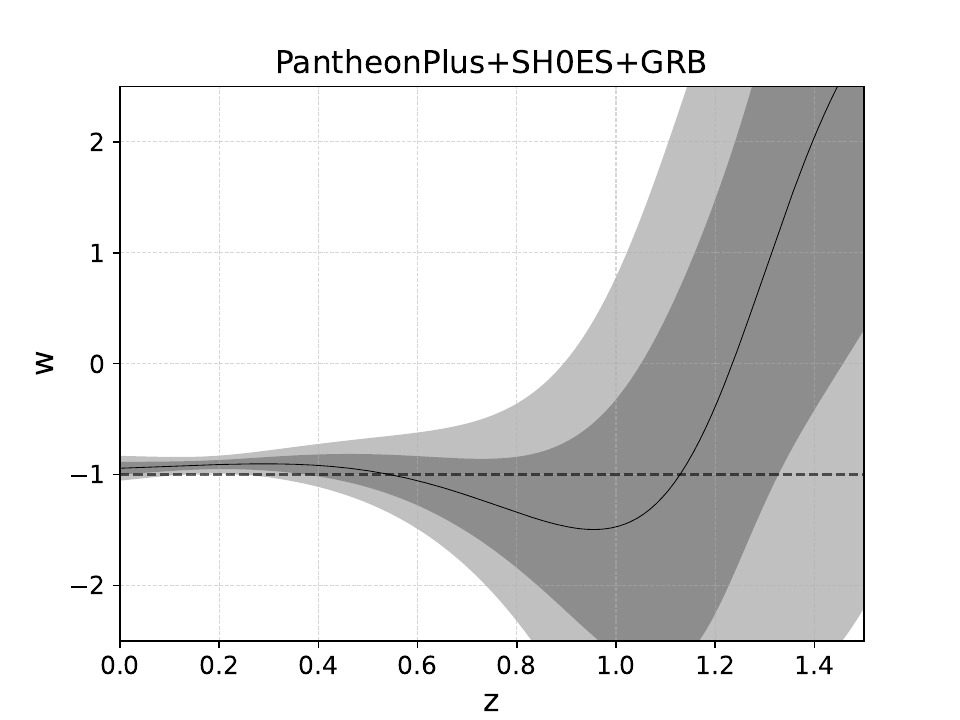}}\\
  \subfloat{\includegraphics[width=0.32\linewidth]{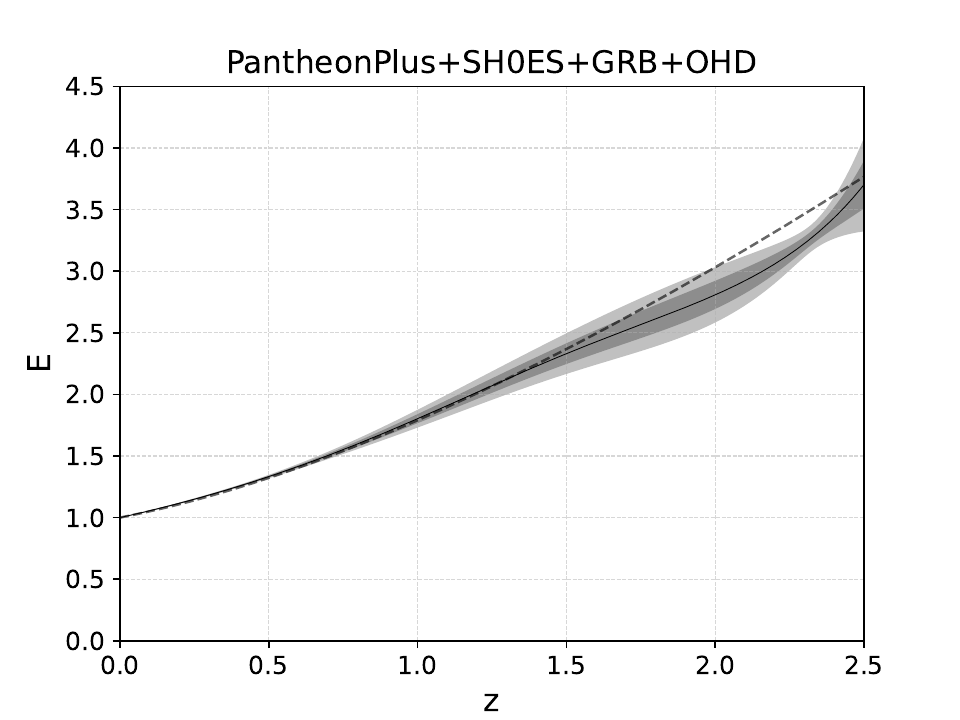}}
  \subfloat{\includegraphics[width=0.32\linewidth]{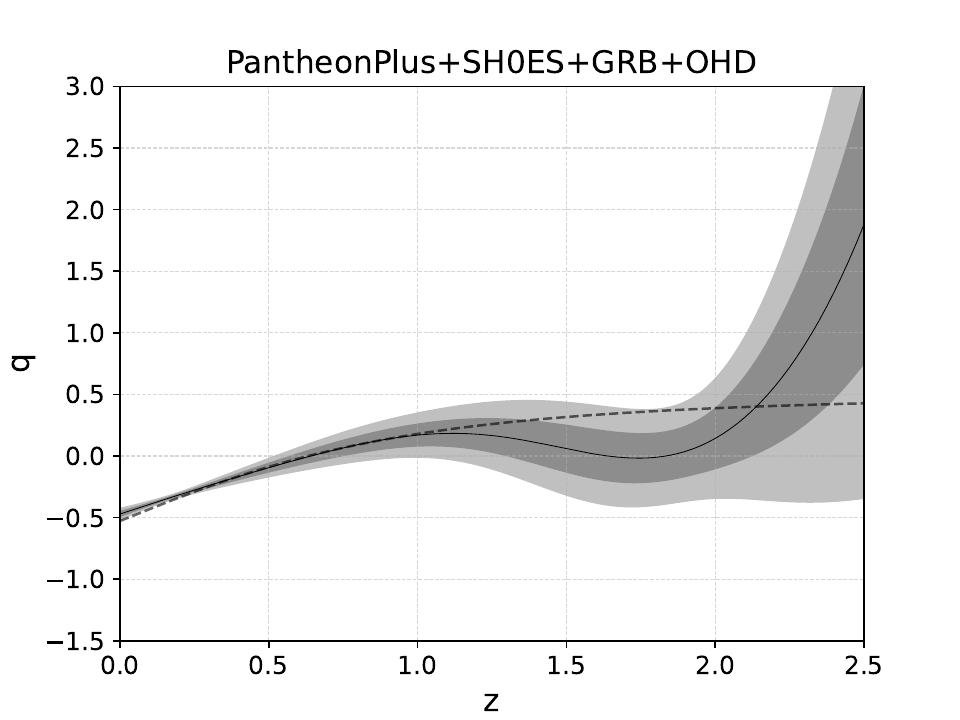}} 
  \subfloat{\includegraphics[width=0.32\linewidth]{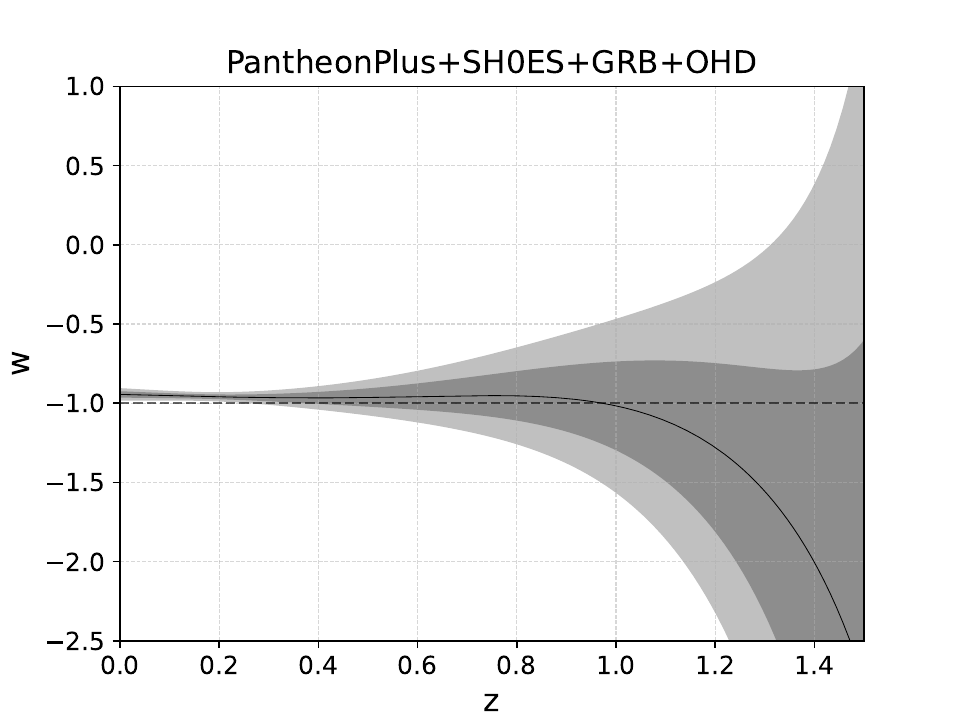}}\\
  \subfloat{\includegraphics[width=0.32\linewidth]{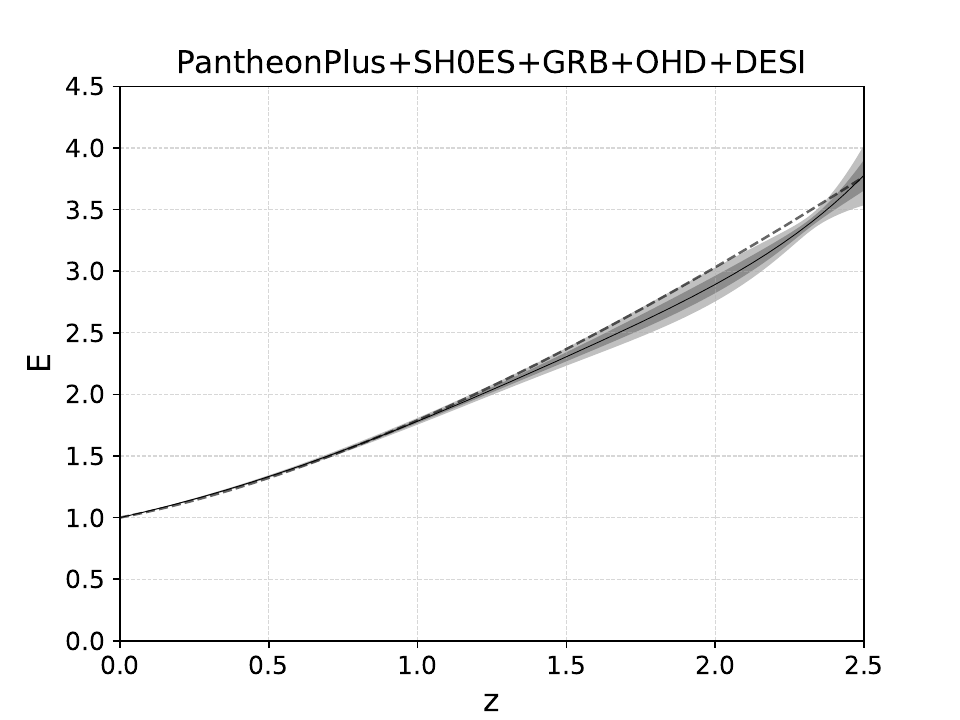}}
  \subfloat{\includegraphics[width=0.32\linewidth]{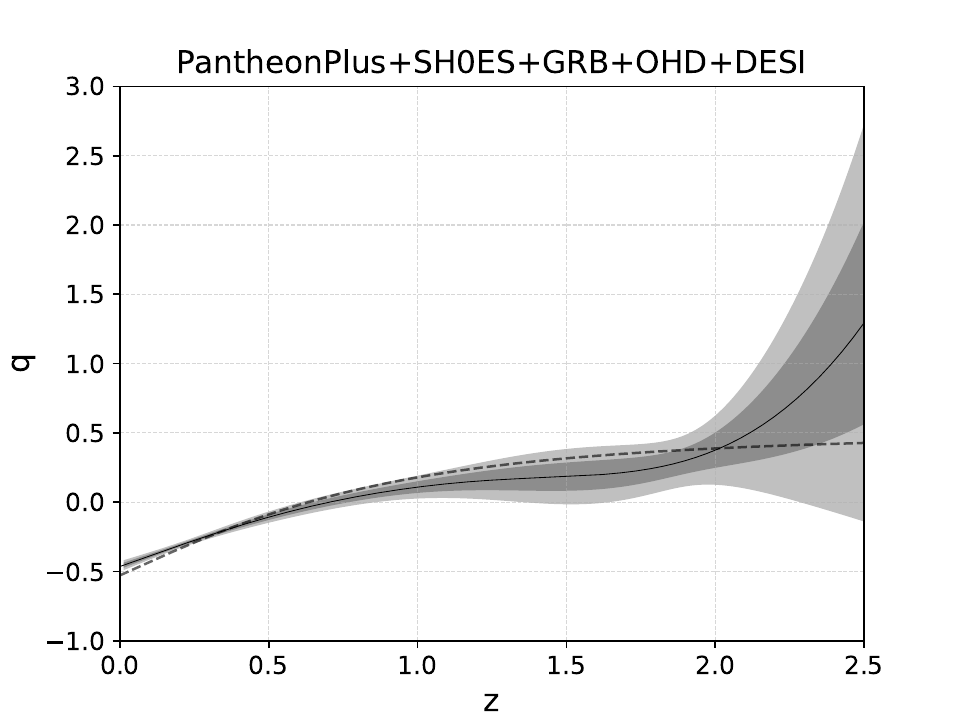}}
  \subfloat{\includegraphics[width=0.32\linewidth]{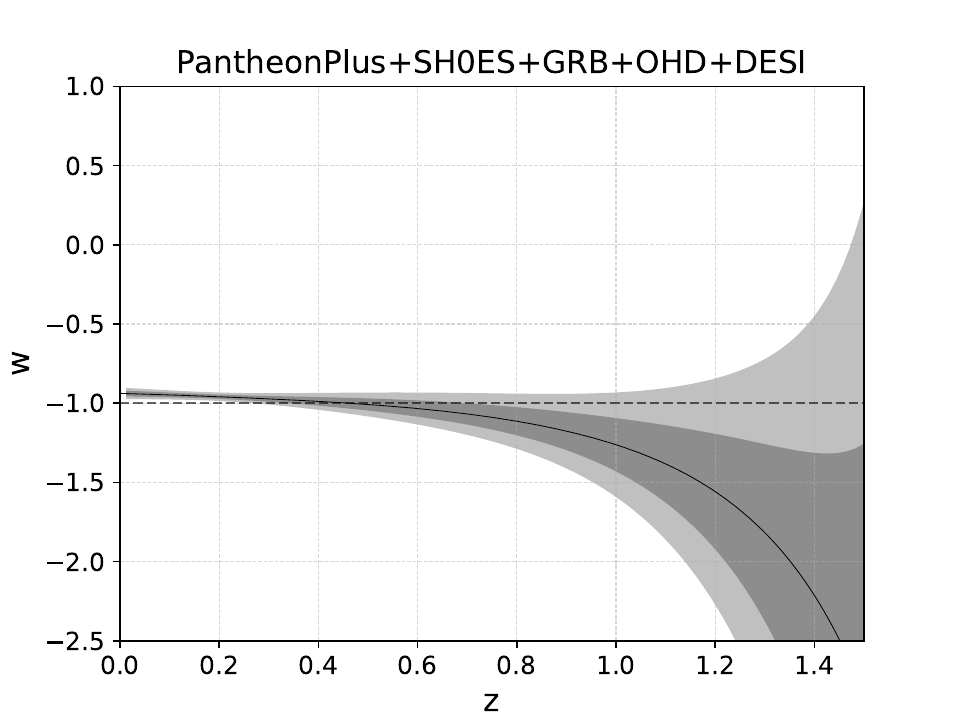}}
  \caption{Reconstruction of the evolution of $E$, $q$, and $w$ utilizing various observation datasets. The observation data employed for these four rows correspond to PantheonPlus+SH0ES, PantheonPlus+SH0ES+GRB, PantheonPlus+SH0ES+GRB+OHD, and PantheonPlus+SH0ES+GRB+OHD+DESI, respectively. The shaded areas in dark grey and light grey are the confidence intervals of 1$\sigma$ and 2$\sigma$, respectively. The black dashed line represents the theoretical curve of the $\Lambda$CDM model. }
  \label{fig_Eqw}
\end{figure*}

We present the results of reconstructing $E$ from various joint datasets in the first column of Fig. \ref{fig_Eqw}. It is evident that $E(z=0)=1$ appears in the reconstruction results, which can be attributed to our choice of using $D^{'}(z=0)=1$ as the initial condition. From the first row of Fig. \ref{fig_Eqw}, we observe that the result reconstruction of PantheonPlus+SH0ES data for $E$ fits well with the $\Lambda$CDM model. However, it exhibits a significant error at high redshift, characterized by an unstable steep increase followed by a steep decrease in the reconstructed mean value. Upon incorporating GRB data into our analysis, we found that the reconstruction error has been reduced, but an instability still persisted with sharp increase and decrease at high redshifts.

The inclusion of OHD and DESI data significantly improved the reconstruction results for $E$. This improvement correlates with a marked reduction in error following $D^{'}$ reconstruction when OHD and DESI data were added. We also observed deviations from the $\Lambda$CDM model within specific ranges: for PantheonPlus+SH0ES+GRB+OHD reconstruction of $E$, deviations occurred within $z\in(1.97, 2.43)$, while for PantheonPlus+SH0ES+GRB+OHD+DESI, deviations were noted within $z\in(2.15, 2.37)$. As illustrated in Fig.~\ref{fig_D}, OHD deviate from the $\Lambda$CDM model within these ranges, thereby influencing both reconstruction results of $D^{'}$ and subsequently those of $E$.

\subsection{The reconstruction of $q$}
\label{subsec:rec q}

The second column of Fig.~\ref{fig_Eqw}  illustrates the deceleration parameter $q$ reconstructed using various joint datasets. The sign of the deceleration parameter, whether positive or negative, indicates whether the universe is undergoing decelerating or accelerating expansion. Consequently, the position of $z~(q=0)$ signifies the point at which the expansion of the universe transitions from acceleration to deceleration or vice versa. We refer to this special redshift as transition redshift and represent it as $z_{\rm t}$. The transition redshift derived from different joint data is summarized in Table \ref{tab:zt}. It is evident that the reconstruction results for four joint datasets exhibit similar transitional redshifts within low redshift ranges ($0.5<z<1.5$). Furthermore, with an increasing amount of data incorporated into these analyses, the value of the transition redshift tends to converge towards the theoretical value predicted by the $\Lambda$CDM model.

\begin{table}[!htbp]
\caption{The transition redshift at which the expansion of Universe shifts from a decelerating phase to an accelerating one (i.e. the deceleration parameter $q$ crosses 0). The error represents the $1 \sigma$ confidence interval.}
\label{tab:zt}       
\begin{tabular}{ll}
\hline\noalign{\smallskip}
datasets & $z_{\rm t}$ \\
\noalign{\smallskip}\hline\noalign{\smallskip}
PantheonPlus+SH0ES & $1.200^{+0.301}_{-0.228}$ \\
PantheonPlus+SH0ES+GRB & $0.911^{+0.277}_{-0.291}$ \\
PantheonPlus+SH0ES+GRB+OHD & $0.631^{+0.108}_{-0.065}$ \\
PantheonPlus+SH0ES+GRB+OHD+DESI & $0.702^{+0.063}_{-0.052}$ \\
\noalign{\smallskip}\hline
\end{tabular}
\end{table}

At low redshift, the results of reconstructing $q$ from joint data are well consistent with the $\Lambda$CDM model, with the exception of the results derived from PantheonPlus+SH0ES +GRB+OHD and PantheonPlus+ SH0ES+GRB+OHD+DESI. These latter results deviate slightly from the curve of $\Lambda$CDM model within the 2$\sigma$ confidence range for $z<0.3$. This discrepancy may be attributed to our exclusion of the data point at $z=0.295$ in the DESI DR2 dataset. If there is a way to incorporate this data point to the DESI DR2 dataset, it could potentially enhance our results at low redshift.

\subsection{The reconstruction of $w$}
\label{subsec:rec w}

From the first two rows of the third column in Fig.~\ref{fig_Eqw}, we observe that the reconstructions of $w$ exhibit a phenomenon characterized by sharp increases and decreases. We propose that this behavior may be attributed not only to the observed data itself but also to the underlying model. Since $w$ is represented by Eq.~\eqref{5} in the $\Lambda$CDM model, even a slight change in the numerator can become significantly amplified when the denominator approaches 0, ultimately resulting in substantial fluctuations in the value of $w$. The results indicate that the reconstruction of $w$ using PantheonPlus+SH0ES and PantheonPlus+SH0ES +GRB is consistent with the $\Lambda$CDM model within the 2$\sigma$ confidence interval, although there are notable differences between the two results. At $z=0$, the PantheonPlus+SH0ES reconstruction yields $w<-1$, while the PantheonPlus+ SH0ES+GRB reconstruction results in $w>-1$.

After incorporating OHD and DESI data, we observed improved constraints in the reconstructed results. However, in contrast to the results from PantheonPlus+SH0ES and PantheonPlus+SH0ES+GRB reconstruction, both sets of data reconstruction deviated from the $\Lambda$CDM model curve within the range of $z<0.3$. From the third and fourth rows of the third column in Fig.~\ref{fig_Eqw}, we observed the evolutionary behavior transitioning from $w<-1$ to $w>-1$. This suggests a potential presence of dynamic dark energy in the late universe. We use $z_{\rm wt}$ to denote the transition redshift where $z~(w=-1)$. 
From the reconstruction results of PantheonPlus+ SH0ES+GRB+OHD+DESI, the value of $z_{\rm wt}$ is around $z_{\rm wt}=0.464^{+0.235}_{-0.120}$. This evolutionary trend, along with the value of $z_{\rm wt}$, is consistent well with the conclusions presented in Ref.~\cite{Pang:2024qyh}, where a parameterized $w(z)$ of dark energy with three redshift bins, were constrained using  Markov-chain Monte Carlo (MCMC) method.
Our results for the parameter $w$ obtained from PantheonPlus+SH0ES+GRB+OHD+DESI are compared with those of the $w_0 w_a$ model presented in Table V of the DESI DR2 \cite{DESI:2025zgx}. Our findings align with the scenario involving DESI+CMB+Pantheon+. Currently, the value of $w$ that we have derived falls within the range of $(-0.8,-1)$.

\section{Discussion and Conclusion}
\label{sec:d and c}

In this article, we employ model-independent method utilizing Gaussian processes to reconstruct the dimensionless Hubble parameter $E$, the deceleration parameter $q$, and the equation of state for dark energy $w$. Regarding the data used, we incorporate a combination of Type Ia supernovae, gamma ray bursts, observational $H(z)$ data, and BAO measurements from DESI DR2.

Overall, PantheonPlus+SH0ES and PantheonPlus +SH0ES+GRB have enhanced the constraints on Gaussian process reconstruction at low redshifts  compared to Union2.1 \cite{Seikel:2012uu}, JLA \cite{Wang:2019ufm} and Pantheon \cite{Lin:2019cuy}. However, their constraints at high redshifts remain insufficient. To address this issue, we employ GRB data with a high redshift. Then we not only combined the OHD, but also added the BAO data provided by DESI. We have obtained the most accurate reconstruction results of $D$ and $D^{'}$ to date. However, as illustrated in Fig. \ref{fig_D}, obtaining an accurate reconstruction of $D^{''}$ continues to present a significant challenge. challenging. From Eqs. (\ref{3}), (\ref{5}) and (\ref{6}), it is evident that the influence of higher-order derivative of $D$ gradually increases on the values of $E$, $q$, and $w$. This explains why the reconstruction error for $E$ is the minimal, followed by that for $q$, while the error for $w$ is the largest.

The result of reconstructing $E$ using PantheonPlus +SH0ES and PantheonPlus+SH0ES+GRB has an abnormal steep increase and then steep decrease around $z=2.5$. This is due to the fact that the value of $D^{'}$ in the reconstruction result is close to 0 near $z=2.5$. Since the reciprocal of $D^{'}$ is $E$, we observe that the value of $E$ experiences a sharp increase as $D^{'}$ approaches 0. Additionally, we noticed that the reconstruction of $E$ by PantheonPlus+SH0ES and PantheonPlus+SH0ES+GRB yield $D^{'}<0$ at $z=2.5$. This situation implies Hubble parameters $H<0$, which lacks physical significance. Therefore, the reconstruction results obtained from PantheonPlus+SH0ES and PantheonPlus+SH0ES+GRB are unreliable at high redshifts.
The reconstruction results of $E$ indicate that incorporating OHD and DESI BAO data as constraints on the derivative of the reconstruction function can significantly enhance the accuracy of the reconstruction results. We observed that the reconstructions of $E$ using PantheonPlus+SH0ES+GRB+OHD and PantheonPlus+SH0ES+GRB+OHD+DESI show a deviation from the $\Lambda$CDM model within a confidence range of 2$\sigma$ at $z>2$. This discrepancy is attributed to the deviation of OHD data from the $\Lambda$CDM model in the redshift range of $z>2$.

The results of reconstructing $q$ using PantheonPlus+ SH0ES and PantheonPlus+SH0ES+GRB are consistent with the $\Lambda$CDM model within 2$\sigma$. The incorporation of $H(z)$ data effectively reduces the uncertainty in the reconstruction results. The reconstruction results of PantheonPlus+SH0ES+GRB+OHD and PantheonPlus+SH0ES+GRB+OHD+DESI provide support for the phenomenon of accelerated expansion in the late universe, revealing no indication of deceleration.
Our results deviate from the curve of $\Lambda$CDM model within the 2$\sigma$ confidence range for $z<0.3$.
The results of PantheonPlus+SH0ES+GRB+OHD and PantheonPlus + SH0ES+GRB+OHD+DESI reconstruction of $w$ suggest the possible existence of dynamic dark energy. The results indicate an evolutionary behavior transitioning from $w<-1$ to $w>-1$ around $z_{\rm wt}=0.464^{+0.235}_{-0.120}$.

\ 

\textit{Acknowledgements}
This work is supported by the National Natural Science Foundation of China (Grant Nos. 12175192, 12005183 and 12005184).

\bibliography{ref.bib}

\end{document}